\documentclass[twocolumn,showpacs,preprintnumbers,amsmath,amssymb]{revtex4-1}
\usepackage{graphicx}% Include figure files
\usepackage{dcolumn}% Align table columns on decimal point
\usepackage{bm}% bold math
\usepackage[compact]{titlesec}
\usepackage{hyperref}
\newcommand\bmtimes{{\bm{\times}}}
\newcommand\bmcdot{{\bm{\cdot}}}

\newcommand\ret{{\mathrm{ret}}}

\newcommand\To{T_{\scriptscriptstyle{{(0)}}}}
\newcommand\brTo{\brT_{\scriptscriptstyle{{(0)}}}}

\newcommand{\half}{{{\textstyle\frac{1}{2}}}}
\newcommand{\quarter}{{{\textstyle\frac{1}{4}}}}
\newcommand{\be}{\begin{equation}}
\newcommand{\ee}{\end{equation} }
\newcommand{\beqa}{\begin{eqnarray} }
\newcommand{\eeqa}{\end{eqnarray} }
\newcommand{\ba}{\begin{array}}
\newcommand{\ea}{\end{array}}
\newcommand{\bpm}{\begin{pmatrix}}
\newcommand{\epm}{\end{pmatrix}}

\newcommand{\ODD}{\mathbf{O}(D,D)}

\newcommand\Phis{\Phi_{{\bfs}}}
\newcommand\EK{K}

\newcommand\rd{{\rm d}}

\newcommand\bfx{{\mathbf{x}}{}}
\newcommand\bfs{{\mathbf{s}}{}}

\newcommand\bfB{{\mathbf{B}}}

\newcommand\bfH{{\mathbf{H}}}
\newcommand\bfK{{\mathbf{K}}}

\newcommand\bfS{{\mathbf{S}}}

\newcommand\eff{{\rm{\bf{eff.}}}}
\newcommand\rhoeff{{\rho_{\eff}}\!}

\newcommand\cH{{\cal H}}

\newcommand\bfhn{{\mathbf{\hat{n}}}}

\newcommand\whK{{\widehat{K}}}
\newcommand\whT{{\widehat{T}}}

\newcommand\dis{\displaystyle}

\def\brh{\bar{h}}

\def\brB{\bar{B}}

\def\brH{\bar{H}}
\def\brK{\bar{K}}

\def\brT{{\bar{T}}}

\newcommand{\brphi}{\bar{\phi}}

\newcommand{\na}{{\nabla}}
\newcommand{\bmna}{\bm{\nabla}}
\newcommand{\trd}{{\bigtriangledown}}

\begin{document}

%\preprint{APS/123-QED}
\title{Stringy Newton Gravity with $H$-flux}
\author{Kyoungho Cho}
\email{khcho23@sogang.ac.kr}
\author{Kevin Morand}
\email{morand@sogang.ac.kr}
\author{Jeong-Hyuck Park\,}
\email{park@sogang.ac.kr}

\affiliation{\vspace{1pt}Department of Physics, Sogang University, 35 Baekbeom-ro, Mapo-gu, Seoul 04107,  Korea}

\begin{abstract}
\centering\begin{minipage}{\dimexpr\paperwidth-6.7cm}
\noindent  A Symmetry Principle   has  been shown to   augment  unambiguously  the  Einstein Field Equations,  promoting         the whole    closed-string massless NS-NS sector  to stringy graviton fields. Here we consider   its  weak field approximation,  take a non-relativistic limit, and   derive the stringy   augmentation of  Newton Gravity:  
\vspace{-12pt}
\[
\begin{array}{lll}
\bm{\nabla}^{2}\Phi=4\pi G \rho+\mathbf{H}{\bm{\cdot}}\mathbf{H}\,,
\quad&\qquad\bm{\nabla}{\bm{\cdot}}\mathbf{H}=0\,,
\quad&\qquad
\bm{\nabla}\bm{\times}\mathbf{H}=4\pi G\, \mathbf{K}\,.
\end{array}\vspace{-7pt}
\]
Not only the mass density~$\rho$ but also the   current density $\mathbf{K}$ is  intrinsic to matter.  Sourcing  $\mathbf{H}$ which is of NS-NS $H$-flux origin,   $\mathbf{K}$ is    nontrivial  if the matter is   `stringy'. $\mathbf{H}$ contributes quadratically  to the Newton potential, but otherwise is decoupled from  the  point particle dynamics, \textit{i.e.~}$\mathbf{\ddot{x}}=-\bm{\nabla}\Phi$. 
We define `stringization'  analogous to magnetization  and  discuss regular as well as  monopole-like singular solutions. 
\end{minipage}
\end{abstract}

%\pacs{11.25.-w}% PACS, the Physics and Astronomy
                             
%%%
%%11.25.-w 	Strings and branes                             
\maketitle
\section{Introduction}
\vspace{5pt}
One of the fundamental  problems  in physics today  is the dark matter problem. Despite  a variety of   observational indications,  \textit{e.g.~}galaxy rotation curves~\cite{Rubin:1980zd},  the Bullet Cluster (1E0657-558)~\cite{Clowe:2006eq}, and ghostly galaxies without  dark matter~\cite{ghost},    no single experiment  has ever  succeeded  in the direct detection of  dark matter particles.    Alternative hypotheses,  including notably  MOND~\cite{Milgrom:1983ca},  pass over the    Equivalence Principle  and   modify General Relativity (GR),     but only give  partial  explanations while  being often   accused  of harming the mathematical beauty thereof.% from an aesthetic perspective.

Developments   over the last decade in an area of string theory, now called  Double Field Theory (DFT),  have gradually  unveiled  a new  form of pure gravity~\cite{Siegel:1993xq,Siegel:1993th,Hull:2009mi,Hull:2009zb,Hohm:2010jy,
Hohm:2010pp,Jeon:2010rw,Jeon:2011cn,Park:2013mpa,Park:2015bza,
Angus:2018mep}. It is based on the  $\ODD$ Symmetry Principle, with   $D$ denoting   the spacetime dimension.  The symmetry can be broken only spontaneously but never explicitly, as the theory is constructed in terms of strictly  $\ODD$-covariant  field variables, namely   the DFT-dilaton~$d$ and DFT-metric~$\cH_{AB}$ (or more powerfully DFT-vielbeins),  forming  the new pure gravity sector.  The Symmetry Principle   further fixes    their  coupling  to generic matter contents which should   also be  in $\ODD$ representations. Examples include  Yang--Mills~\cite{Jeon:2011kp,Hohm:2011ex},  fermions~\cite{Jeon:2011vx} (\textit{c.f.~}\cite{Coimbra:2011nw}), R-R sector~\cite{Rocen:2010bk,Hohm:2011zr,Hohm:2011dv,Jeon:2012kd}, full-order supersymmetrizations~\cite{Jeon:2011sq,Jeon:2012hp}, point particles~\cite{Ko:2016dxa,Blair:2017gwn},
fundamental strings~\cite{Hull:2006va,Hull:2009sg,Lee:2013hma,
Blair:2013noa,Park:2016sbw,Blair:2019qwi}, and the Standard Model itself~\cite{Choi:2015bga}. Naturally,  the Einstein Field Equations are   augmented to an $\ODD$-symmetric form~\cite{Angus:2018mep} (\textit{c.f.~}\cite{Park:2019hbc} for a short summary):
\be
G_{AB}={\frac{8\pi G}{c^{4}}} T_{AB}\,,
\label{MASTER}
\ee
which carry $\ODD$ vector indices and unify the Euler--Lagrange equations of all  the stringy graviton fields, 
$\left\{\cH_{AB},d\right\}$.  In a parallel manner to GR,  $G_{AB}$ is the off-shell conserved $\ODD$-symmetric Einstein curvature   constructed out of $\left\{\cH_{AB},d\right\}$~\cite{Park:2015bza}, while $T_{AB}$ is the on-shell conserved energy-momentum  tensor, defined through the variation of the matter Lagrangian with respect to 
$\left\{\cH_{AB},d\right\}$~\cite{Angus:2018mep}.

Remarkably, the perfectly $\ODD$-symmetric vacua, satisfying ${G_{AB}=0}$, turned out to  be a  topological phase  which allows    no moduli and  no  interpretation within   Riemannian geometry, thus  escaping   beyond the realm of GR~\cite{Lee:2013hma,Morand:2017fnv,Cho:2018alk}. Only after a spontaneous symmetry breaking of $\ODD$, the familiar string theory backgrounds characterized  by the Riemannian metric $g_{\mu\nu}$ and the Kalb--Ramond skew-symmetric two-form  potential $B_{\mu\nu}$ emerge:  these component fields  parametrize  the DFT-metric while  being identified   as the Nambu--Goldstone bosons~\cite{Berman:2019izh}. The master formula~(\ref{MASTER}) then 
reduces to  (\textit{c.f.~}\cite{Cho:2019ofr} for non-Riemannian cases)
\vspace{-5pt}
\be
\ba{r}
\half e^{2\phi}\trd^{\rho}\!\left(e^{-2\phi}H_{\rho\mu\nu}\right)=\,\frac{8\pi G}{c^{4}}\EK_{[\mu\nu]}\,,\\
R_{\mu\nu}+2\trd_{\mu}(\partial_{\nu}\phi)-\quarter H_{\mu\rho\sigma}H_{\nu}{}^{\rho\sigma}
=\frac{8\pi G}{c^{4}}\EK_{(\mu\nu)}\,,\\
R+4\trd_{\mu}(\partial^{\mu}\phi)-4\partial_{\mu}\phi\partial^{\mu}\phi-\textstyle{\frac{1}{12}}H_{\lambda\mu\nu}H^{\lambda\mu\nu}=\,\frac{8\pi G}{c^{4}}\To\,,~~
\ea
\label{REDFE}
\ee
which imply a pair of conservation laws, \vspace{-5pt} 
\be
\ba{r}
\na^{\mu}\!\left(e^{-2\phi}\EK_{[\mu\nu]}\right)=0\,,\\
\na^{\mu}\EK_{(\mu\nu)}-2\partial^{\mu}\phi\,\EK_{(\mu\nu)}+\half H_{\nu}{}^{\lambda\mu}\EK_{[\lambda\mu]}-\half\partial_{\nu}\To=0\,.
\ea
\label{CON}
\ee
The left and the right  hand sides of the equalities in (\ref{REDFE})  come from $G_{AB}$ and  $T_{AB}$ in (\ref{MASTER}) separately. Schematically, $K_{[\mu\nu]}$, $K_{(\mu\nu)}$, and $\To$  are   the energy-momentum tensor   components relevant to $B_{\mu\nu}$,  $g_{\mu\nu}$,  and the string dilaton $\phi=d+\quarter\ln\left|g\right|$, respectively.  Having said that,  since (\ref{MASTER}) is derived from the variations of the $\ODD$ covariant fields, in particular the DFT-dilaton $d$ rather than the string dilaton $\phi$, the  specific form of the  Einstein curvature, $R_{\mu\nu}-\half g_{\mu\nu}R$,  does not automatically appear  in (\ref{REDFE}).

We stress that for each  matter content, the $\ODD$ symmetry  determines   its  coupling to the closed-string massless NS-NS sector of $\{B_{\mu\nu},g_{\mu\nu},\phi\}$ and hence fixes  $K_{\mu\nu},\To$ completely~\cite{Angus:2018mep}.  For example,  a point particle should couple to the string frame metric~$g_{\mu\nu}$ only   ---minimally in the standard way--- resulting in  $K_{[\mu\nu]}=0$, $\To=0$, and 
\be
\dis{
K^{\mu\nu}(x)=\frac{mc}{2}\int\rd\tau~\dot{y}^{\mu}(\tau)\dot{y}^{\nu}(\tau)\frac{\delta\big(x-y(\tau)\big)e^{2\phi}}{\sqrt{-g}}\,.}
\label{Kparticle}
\ee
The particle  follows    geodesics defined in the string frame,
\be
\ddot{x}^{\lambda}+\gamma^{\lambda}_{\mu\nu}\dot{x}^{\mu}\dot{x}^{\nu}=0\,.
\label{geodesic}
\ee
Rewriting this  in the Einstein frame would involve  the gradient of $\phi$ and thus obscure the Equivalence Principle. That is to say, the $\ODD$ Symmetry Principle asserts  that  the Equivalence Principle  for a point particle should  hold    in the string frame. On the other hand, fundamental strings, spinorial fermions, and  the R-R sector couple  to $B_{\mu\nu}$ and $g_{\mu\nu}$ but not to  $\phi$, resulting in  asymmetric $K_{\mu\nu}$ and (still) trivial $\To$.  Further,  in contrast,  gauge bosons   couple to  $g$ and $\phi$, but do not interact with the $B$-field~(\ref{photon})\,\cite{Choi:2015bga}. While the $B$-field  does not interact with the electromagnetic force nor point particles,  its electric $H$-flux  nevertheless contributes to the mass formula~\cite{Angus:2018mep},
\be
M=\int e^{-2d}\left|\,2K_{0}{}^{0}+\frac{1}{16\pi G}H_{0\mu\nu}H^{0\mu\nu}\,\right|\,.
\label{MASS}
\ee
In fact, having a larger profile than the (visible) matter represented  here by $K_{0}{}^{0}$, the electric $H$-flux can produce non-monotonic non-Keplerian rotation curves~\cite{Ko:2016dxa}.  In this way,   $H$-flux  behaves  like   dark matter~\cite{Park:2017snt}, \textit{c.f.~}\cite{Cheung:2008fz}.  Note that its dual scalar  is also known as  a dark matter candidate,  `an axion'~\cite{Kim:1979if,Shifman:1979if,Dine:1981rt,Zhitnitsky:1980tq,Svrcek:2006yi}.

It is the purpose of the present paper to consider the weak field approximation of (\ref{REDFE}), (\ref{CON}), (\ref{geodesic}) for  ${D=4}$,   take a consistent non-relativistic limit,  and  faithfully derive the stringy augmentation of Newton Gravity   spelled out  in the Abstract. We hope our work may deepen  the  physical   understanding of    the $\ODD$-completed General Relativity~(\ref{MASTER}) and contribute to   examining rigorously   the prospect of  $H$-flux as a dark matter candidate.  

%%%%%%%%%%%%%%%%%%%%%%%%%%%%%%%%%%%%%%%%%%%%%%%%%%%
%%%%%%%%%%%%%%%%%%%%%%%%%%%%%%%%%%%%%%%%%%%%%%%%%%%

\vspace{5pt}
\section{Lorentz symmetric weak field approximation\label{SEClEDFE}}
\vspace{5pt}
We  start our  weak field approximation of (\ref{REDFE}) by linearizing  the metric, 
$g_{\mu\nu}=\eta_{\mu\nu}+h_{\mu\nu}$, $g^{\mu\nu}\simeq\eta^{\mu\nu}-h^{\mu\nu}$, around a flat Minkowskian background  with trivial $H$-flux and  dilaton~$\phi$. The spacetime indices are raised or lowered by the constant   metric~$\eta$, \textit{e.g.~}$h^{\mu\nu}=\eta^{\mu\rho}\eta^{\nu\sigma}h_{\rho\sigma}$.
The linearized  Riemann curvature,
\be
R^{\kappa}{}_{\lambda\mu\nu}=\half\left(
\partial_{\lambda}\partial_{\mu}h^{\kappa}{}_{\nu}
-\partial_{\lambda}\partial_{\nu}h^{\kappa}{}_{\mu}
-\partial^{\kappa}\partial_{\mu}h_{\lambda\nu}
+\partial^{\kappa}\partial_{\nu}h_{\lambda\mu}\right)\,,\vspace{-3pt}
\label{lR}
\ee
is invariant under the linearized diffeomorphisms,\vspace{-8pt}
\be
\ba{ll}
\delta_{\xi} h_{\mu\nu}=\partial_{\mu}\xi_{\nu}+\partial_{\nu}\xi_{\mu}\,,\qquad&\qquad
\delta_{\xi} R^{\kappa}{}_{\lambda\mu\nu}= 0\,.
\ea
\ee
Using $\delta_{\xi}(\partial_{\rho}h^{\rho}{}_{\mu}-\half\partial_{\mu}h^{\rho}{}_{\rho})=\partial_{\rho}\partial^{\rho}\xi_{\mu}$, we fix the gauge,
\be
\partial_{\rho}h^{\rho}{}_{\mu}-\half\partial_{\mu}h^{\rho}{}_{\rho}+2\partial_{\mu}\phi=0\,.
\label{gauge}
\ee
It follows from (\ref{lR}) and (\ref{gauge}),\vspace{-8pt}
\be
\ba{r}
R_{\mu\nu}= -\half\partial_{\rho}\partial^{\rho}h_{\mu\nu}-2\partial_{\mu}\partial_{\nu}\phi\,,~\\
\partial_{\rho}\partial^{\rho}\phi=\quarter \partial_{\rho}\partial^{\rho}h^{\sigma}{}_{\sigma}-\half\partial_{\rho}\partial_{\sigma}h^{\rho\sigma}\,,
\ea
\ee
and further, from the integrability of (\ref{gauge}),
\be
\partial_{\mu}\partial_{\rho}h^{\rho}{}_{\nu}=\partial_{\nu}\partial_{\rho}h^{\rho}{}_{\mu}\,.
\label{integrability}
\ee
Similarly, we choose a  gauge  for the $B$-field,\vspace{-8pt}
\be
\ba{lll}
\delta_{\lambda}B_{\mu\nu}=\partial_{\mu}\lambda_{\nu}-\partial_{\nu}\lambda_{\mu}\quad&\quad\Longrightarrow\quad&\quad
\partial_{\rho}B^{\rho}{}_{\mu}=0\,.
\ea\label{gaugeB}
\ee
Now, assuming the following  scales, 
\be
h_{\mu\nu}\,\sim\, K_{(\mu\nu)}\,\sim\, \phi\,\sim\, \To\,\sim\, \left(H_{\lambda\mu\nu}\right)^{2}\,\sim\,\left(K_{[\mu\nu]}\right)^{2}\,,
\label{lscale}
\ee
the consistent  linearization of (\ref{REDFE})  can be achieved,  \vspace{-8pt}
\be
\ba{rcl}
\partial^{\rho}H_{\rho\mu\nu}\,=\,\partial_{\rho}\partial^{\rho}B_{\mu\nu}&=&\frac{16\pi G}{c^{4}}\EK_{[\mu\nu]}\,,\\
\partial_{\rho}\partial^{\rho}h_{\mu\nu}+\half H_{\mu\rho\sigma}H_{\nu}{}^{\rho\sigma}&=&-\frac{16\pi G}{c^{4}}\EK_{(\mu\nu)}\,,\\
\partial_{\rho}\partial_{\sigma}h^{\rho\sigma}+\textstyle{\frac{1}{12}}H_{\rho\sigma\tau}H^{\rho\sigma\tau}&=&-\frac{8\pi G}{c^{4}}\To\,,
\ea
\label{lEDFE}
\ee
which, with  (\ref{integrability}),  imply   the following linearized conservation equations, \vspace{-8pt}
\be
\ba{ll}
\partial^{\rho}K_{[\rho\mu]}=0\,,\quad&\quad
\partial^{\rho}K_{(\rho\mu)}+\half H_{\mu}{}^{\rho\sigma}K_{[\rho\sigma]}-\half\partial_{\mu}\To=0\,.
\ea
\label{lCON}
\ee
With a well-known  relation,
\be
\!\!\bmna^{2}\frac{1}{\left|\bfx-\bfx^{\prime}\right|}=
-\bmna\bmcdot\left(\frac{\bfx-\bfx^{\prime}}{\,\,\left|\bfx-\bfx^{\prime}\right|^{3}}\right)
=-4\pi\delta(\bfx-\bfx^{\prime})\,,
\label{standard}
\ee
the first formula in (\ref{lEDFE})  is solved by  a  `retarded potential', 
\be
B_{\mu\nu}(x)=-\dis{\frac{4G}{c^{4}}\int\rd^{3}x^{\prime}~\frac{K_{[\mu\nu]}(x^{0}_{\ret}\,,\,\bfx^{\prime})}{\left|\bfx-\bfx^{\prime}\right|}}\,,
\label{Bmunu}
\ee
where $x^{0}_{\ret}:=x^{0}-\left|\bfx-\bfx^{\prime}\right|$ is the   retarded temporal coordinate.  Thanks to $\partial^{\rho}K_{[\rho\mu]}=0$ (\ref{lCON}), up to a surface integral,  the gauge condition~(\ref{gaugeB}) is indeed fulfilled, 
\be
\dis{
\partial^{\mu}B_{\mu\nu}(x)=-\frac{4G}{c^{4}}\int\rd^{3}x^{\prime}~\frac{\left.\,\partial^{\prime\mu}K_{[\mu\nu]}(x^{\prime})\right|_{x^{\prime 0}\rightarrow x^{0}_{\ret}}}{\left|\bfx-\bfx^{\prime}\right|}=0\,.}
\label{gaugeB2}
\ee
The expression for the $H$-flux follows 
\be
H_{\lambda\mu\nu}(x)=-\dis{\frac{4G}{c^{4}}\int\rd^{3}x^{\prime}~\frac{\,\left.3\partial^{\prime}_{[\lambda}K_{\mu\nu]}(x^{\prime})\right|_{x^{\prime 0}\rightarrow x^{0}_{\ret}}}{\left|\bfx-\bfx^{\prime}\right|}}\,.
\label{Hlmn}
\ee
In a similar fashion,  with some shorthand notations,  
\be
\ba{l}
\whK_{(\mu\nu)}:=\frac{16\pi G}{c^{4}}\EK_{(\mu\nu)}+\half H_{\mu\rho\sigma}H_{\nu}{}^{\rho\sigma}\,,\\
\whT:=\frac{8\pi G}{c^{4}}\To+\textstyle{\frac{1}{12}}H_{\rho\sigma\tau}H^{\rho\sigma\tau}\,,
\ea
\label{barred}
\ee  
we solve the second formula in (\ref{lEDFE}), 
\be
h_{\mu\nu}(x)=\dis{\frac{1}{4\pi}\int\rd^{3}x^{\prime}~\frac{\whK_{(\mu\nu)}(x^{0}_{\ret}\,,\,\bfx^{\prime})}{\left|\bfx-\bfx^{\prime}\right|}}\,.
\label{hmunu}
\ee 
From (\ref{lEDFE}), (\ref{lCON}), and the $H$-flux Bianchi identity,  we get
\be
\dis{\partial^{\mu}\whK_{(\mu\nu)}=\partial_{\nu}\whT}\,,
\ee
which gives,  comparable  to (\ref{gaugeB2}),  
\be
\partial_{\rho}h^{\rho}{}_{\mu}(x)
=\dis{\partial_{\mu}\left[\,\frac{1}{4\pi}\int\rd^{3}x^{\prime}~\frac{\whT(x^{0}_{\ret}\,,\,\bfx^{\prime})}{\left|\bfx-\bfx^{\prime}\right|}\,\right]}\,.
\label{pmu}
\ee
This verifies  for consistency  that (\ref{hmunu})  indeed satisfies   the integrability condition~(\ref{integrability}), while    
 the third formula in (\ref{lEDFE}) is automatically fulfilled as $\partial_{\rho}\partial_{\sigma}h^{\rho\sigma}(x)=-\whT(x)$. 
Lastly, from (\ref{gauge}),  the string dilaton is fixed, 
\be
\phi(x)=\frac{1}{4\pi}\dis{\int\rd^{3}x^{\prime}~
\left.\frac{\,\frac{1}{4}\whK_{\mu}{}^{\mu}(x^{\prime})-\half\whT(x^{\prime})\,}
{\left|\bfx-\bfx^{\prime}\right|}\right|_{x^{\prime 0}\,\rightarrow\, x^{0}_{\ret}}}\,.
\label{phifixed}
\ee
To summarize, through (\ref{Hlmn}), (\ref{hmunu}), and (\ref{phifixed}),   $\{K_{\mu\nu},\To\}$ (`matter') determines $\{H_{\lambda\mu\nu},h_{\mu\nu},\phi\}$ (`geometry').   While $H_{\lambda\mu\nu}$ is given by a single volume integral,  the other two, $h_{\mu\nu},\phi$, involve triple volume integrals  of `matter'.  It is  worth while to note that the  quantity inside the bracket in (\ref{pmu}) is related to the $\ODD$ singlet integral measure of DFT (scalar density with weight one),
\be
e^{-2d}\!=\!\sqrt{-g}e^{-2\phi}\simeq 1+\half h_{\mu}{}^{\mu}-2\phi=1+\textstyle{\frac{1}{4\pi}
}\dis{\int}\textstyle{\!\rd^{3}x^{\prime}\,\frac{\whT(x^{0}_{\ret}\,,\,\bfx^{\prime})}{\left|\bfx-\bfx^{\prime}\right|}\,.}
\label{em2d}
\ee
The linearized geodesic equation assumes  the form,
\be
\ddot{x}^{\lambda}+\half\!\left(\partial_{\mu}h^{\lambda}{}_{\nu}+\partial_{\nu}h^{\lambda}{}_{\mu}-\partial^{\lambda}h_{\mu\nu}\right)\dot{x}^{\mu}\dot{x}^{\nu}=0\,.
\label{lgeodesic}
\ee

%%%%%%%%%%%%%%%%%%%%%%%%%%%%%%%%%%%%%%%
%%%%%%%%%%%%%%%%%%%%%%%%%%%%%%%%%%%%%%%
%\vspace{5pt}
\section{Non-relativistic limit: Stringy  Newton \label{SECNONR}}
\vspace{5pt}
\noindent We proceed to take the non-relativistic,  large $c$ limit of (\ref{lEDFE}). With $x^{0}=ct$, we let all the fields be functions of $(t,\bfx)=(t, x^{i})$, $i=1,2,3$. We   suppress   $\partial_{0}{\,=\frac{1}{c}\frac{\partial~}{\partial t}}{\,= 0}$ and put  $\partial_{\rho}\partial^{\rho}{\,= \partial_{i}\partial^{i}}{\,=\bmna^{2}}$. Further, for the point particle of (\ref{Kparticle}), (\ref{geodesic}), we set  ${t=\tau}$ (proper time)   and $\left|\dot{\bfx}\right|<<c$, such that the linearized geodesic equation~(\ref{lgeodesic}) reduces  to $\ddot{x}^{i}=\half \partial_{i}h_{00}c^{2}$ only.  Combined with (\ref{gauge}),  (\ref{lEDFE}),  this  fixes all the scales, which are  compatible  with  (\ref{lscale}): \vspace{-7pt}
\be
\ba{rlrl}
h_{\mu\nu}&\!=c^{-2}\brh_{\mu\nu}\,,\quad&\qquad  K_{(\mu\nu)}&\!=c^{2}\brK_{(\mu\nu)}\,,\\
\phi&\!=c^{-2}\brphi\,,\quad&\qquad \To&\!=c^{2}\brTo\,,\\
B_{\mu\nu}&\!=c^{-1}\brB_{\mu\nu}\,,\quad&\qquad K_{[\mu\nu]}&\!=c^{3}\brK_{[\mu\nu]}\,,\\
H_{\lambda\mu\nu}&\!=c^{-1}\brH_{\lambda\mu\nu}\,.\quad&\quad{}&{}
\ea
\label{nrscales}
\ee
This then    straightforwardly  produces  the non-relativistic limit of the previous  weak field approximation:  with  ${\partial_{0}=0}$ assumed, after removing the symbol~$c$, replacing   $x^{0}_{\ret}$ by $t$,   and putting  bars over  each  quantity,  all the formulas  from section~\ref{SEClEDFE}  survive to preserve    their forms and set of  relations.      For example, (\ref{lEDFE}) reduces  to\vspace{-5pt}
\be
\ba{rcl}
\partial^{i}\brH_{i\mu\nu}\,=\,\bmna^{2}\brB_{\mu\nu}&=&16\pi G\brK_{[\mu\nu]}\,,\\
\bmna^{2}\brh_{\mu\nu}+\half \brH_{\mu\rho\sigma}\brH_{\nu}{}^{\rho\sigma}
&=&-16\pi G\brK_{(\mu\nu)}\,,\\
\partial_{i}\partial_{j}\brh^{ij}+\textstyle{\frac{1}{12}}\brH_{\rho\sigma\tau}\brH^{\rho\sigma\tau}&=&-8\pi G\brTo\,,
\ea
\label{nREDFE}
\ee
%and the gauge  conditions~(\ref{gauge}), (\ref{integrability}), (\ref{gaugeB}) give
%\be
%\ba{lll}
%\partial_{i}\brh^{i}{}_{0}=0\,,\quad&\quad
%\partial_{i}\partial_{k}\brh^{k}{}_{j}=
%\partial_{j}\partial_{k}\brh^{k}{}_{i}\,,\quad&\quad\partial^{i}\brB_{i\mu}=0\,.
%\ea
%\label{nrgauge}
%\ee
%Setting the  scale of   $K_{[\mu\nu]}$ one power higher than  $K_{(\mu\nu)}$ in (\ref{nrscales}), we  are able to keep the stringy property of matter, if any,  and let the $H$-flux survive in the limit. 
%From the consevation relations~(\ref{lCON}), the matter content should meet
%\be
%\ba{cc}
%\multicolumn{2}{c}{
%\partial^{j}\brK_{(ji)}+\half %\brH_{ijk}\brK^{[jk]}+\brH_{ij0}\brK^{[j0]}-\half\partial_{i}\brTo=0\,,}\\
%\partial^{j}\brK_{(j0)}+\half \brH_{0jk}\brK^{[jk]}=0\,,\qquad&\qquad
%\partial^{i}\brK_{[i\mu]}=0\,.
%\ea
%\label{nRCON}
%\ee
and (\ref{Bmunu})  becomes\vspace{-5pt}
\be
\brB_{\mu\nu}(x)=-\dis{4G\int\rd^{3}x^{\prime}~\frac{\brK_{[\mu\nu]}(t\,,\,\bfx^{\prime})}{\left|\bfx-\bfx^{\prime}\right|}}\,.
\label{nrBmunu}
\ee
Furthermore,  the  $H$-flux Bianchi identity reads  now 
\be
\partial_{[i}\brH_{jk]0}=0\,.
\label{nrBianchi}
\ee
Hereafter, we focus on the  Newton potential which is the only quantity directly relevant to the particle dynamics, \vspace{-7pt}
\be
\ba{ll}
\Phi:=-\half c^{2}h_{00}=-\half\brh_{00}\,,\quad&\quad\ddot{\bfx}=-\bmna\Phi\,.
\ea
\ee
We then identify all the quantities which can affect the  Newton potential: namely,  
 the mass density~$\rho$, the stringy current density~$\bfK$, and $B$-field/$H$-flux vectors~$\bfB,\bfH$, as follows\vspace{-5pt}
\be
\ba{l}
\rho:=2\brK_{00}\,,\\
\bfK:=2\sqrt{2}\left(\brK_{[01]},\,\brK_{[02]},\,\brK_{[03]}\right)\,,\\
\bfB:=\frac{1}{\sqrt{2}}\left(\brB_{10},\,\brB_{20},\,\brB_{30}\right)\,,\\
\bfH:=\bmna\bmtimes\bfB=\frac{1}{\sqrt{2}}\left(\brH_{023},\,\brH_{031},\,\brH_{012}\right)\,.
\ea
\ee
Crucially, $\{\rho,\bfK,\Phi,\bfH\}$ forms an `autonomy' of closed  relations: from (\ref{nREDFE}), \vspace{-5pt}
\be
\ba{ll}
\bmna^{2}\Phi=4\pi G \rho+\bfH\bmcdot\bfH\,,\qquad&\qquad
\bmna\bmtimes\bfH=4\pi G\, \bfK\,,
\ea
\label{MAIN}
\ee
and, from (\ref{lCON}), (\ref{nrBianchi}),  $\bfK$ and $\bfH$ are both divergenceless, \vspace{-5pt}
\be
\ba{ll}
\bmna\bmcdot\bfK=0\,,\qquad&\qquad\bmna\bmcdot\bfH=0\,.\vspace{-3pt}
\ea
\label{divergenceless}
\ee 
That is to say, the Newton potential  is fully determined by both  the mass density and the stringy  current density: directly so by $\rho$, and   indirectly  so by $\bfK$ as  mediated through     $\bfH$,
\be
\ba{l}
\dis{
\Phi=-G\!\int\!\rd^{3}x^{\prime}\,\frac{\, \rhoeff(t,\bfx^{\prime})\,}{\left|\bfx-\bfx^{\prime}\right|}\,,}\quad~~~~~\,\,
\rhoeff:=\rho+\textstyle{\frac{1}{4\pi G}}\bfH\bmcdot\bfH\,,\\
\dis{
\bfH=
G\!\int\!\rd^{3}x^{\prime}~\bfK(t,\bfx^{\prime})\bmtimes
\frac{(\bfx-\bfx^{\prime})\,}{\left|\bfx-\bfx^{\prime}\right|^{3}}=\bmna\bmtimes\bfB}\,,\\
\dis{
\bfB=G\!\int\!\rd^{3}x^{\prime}~~
\frac{\bfK(t,\bfx^{\prime})}{\left|\bfx-\bfx^{\prime}\right|}\,,\qquad~\,\,~~~\,~\bmna\bmcdot\bfB=0\,.}
\ea
\label{NewtonPot}
\ee
Although $\bfH$ originates from the electric components of the $H$-flux,  its behaviour is identical to the  magnetic field in classical Magnetostatics  such that  a  ``Biot--Savart law"  holds above. The vector potential $\bfB$  has  NS-NS $B$-field origin, and notation-wise should not  be confused with the magnetic field.    It is instructive to note that $\rho_{\eff}$ is consistent with (\ref{MASS}) at the linearized level (\textit{c.f.}~\cite{Angus:2019bqs}), and involves a double volume integral as
\be
\bfH\bmcdot\bfH=G^{2}\!
\int\!\rd^{3}x^{\prime}\!\int\!\rd^{3}x^{\prime\prime}~
\frac{\mathbf{{det}}{\tiny{\left[\ba{cc}(\bfx-\bfx^{\prime})\bmcdot
(\bfx-\bfx^{\prime\prime})&(\bfx-\bfx^{\prime})\bmcdot\bfK^{\prime\prime}\\
(\bfx-\bfx^{\prime\prime})\bmcdot\bfK^{\prime}&
\bfK^{\prime}\bmcdot\bfK^{\prime\prime}\ea\right]}}}
{\left|\bfx-\bfx^{\prime}\right|^{3}\left|\bfx-\bfx^{\prime\prime}\right|^{3}}\,,
\label{exact}
\ee
where  $\bfK^{\prime}=\bfK(t,\bfx^{\prime}), \,\bfK^{\prime\prime}=\bfK(t,\bfx^{\prime\prime})$.  \vspace{5pt}

In  analogy  to  the  magnetization in electrodynamics, 
 we introduce the notion of \textit{stringization}  for the  stringy  current density $\bfK$ which is divergence free, 
\be
\bfK(t,\bfx)=\bmna\bmtimes\bfs(t,\bfx)\,.
\ee
The corresponding  $\bfB$, $\bfH$ are, from (\ref{standard}), (\ref{NewtonPot})~(\textit{c.f.~}\cite{Jackson:1998nia}),
\be
\ba{lll}
\bfB&=&%\dis{G\!\int\!\rd^{3}x^{\prime}~\frac{\,\bmna^{\prime}\bmtimes\bfs(t,\bfx^{\prime})}{\left|\bfx-\bfx^{\prime}\right|}}\vspace{2pt}\\
%{}&=&
\dis{G\!\int\!\rd^{3}x^{\prime}~\frac{\bfs(t,\bfx^{\prime})\bmtimes(\bfx-\bfx^{\prime})}{\left|\bfx-\bfx^{\prime}\right|^{3}}+G\!\oint\rd\mathbf{A}\bmtimes\frac{\bfs(t,\bfx^{\prime})}{\left|\bfx-\bfx^{\prime}\right|}}\,,\vspace{10pt}\\
\bfH&=&\dis{4\pi G\bfs(t,\bfx)+G\!\int\!\rd^{3}x^{\prime}~\frac{3\bfhn^{\prime}\left(\bfhn^{\prime}\bmcdot\,\bfs(t,\bfx^{\prime})\right)-\bfs(t,\bfx^{\prime})}{\left|\bfx-\bfx^{\prime}\right|^{3}}}\vspace{-3pt}\\
{}&=&4\pi G\bfs(t,\bfx)- G\bmna\Phis(t,\bfx)\,,
\ea
\label{HPhis}
\ee
in which $\bfhn^{\prime}=\frac{\bfx-\bfx^{\prime}}{\left|\bfx-\bfx^{\prime}\right|}$   and $\Phis$ is a \textit{stringy scalar potential},
\be
\dis{
\Phis(t,\bfx)=\!\int\!\rd^{3}x^{\prime}\,\frac{\,\bfs(t,\bfx^{\prime})\bmcdot(\bfx-\bfx^{\prime})}{\left|\bfx-\bfx^{\prime}\right|^{3}}=-\bmna\bmcdot\!\!
\int\!\rd^{3}x^{\prime}\,\frac{\bfs(t,\bfx^{\prime})}{\left|\bfx-\bfx^{\prime}\right|}
\,.
}
\ee
Clearly from (\ref{HPhis}),   $\bmna\bmtimes\bfH=4\pi G\bmna\bmtimes\bfs$.  Far away from a localized  source, $\left|\bfx\right|{>>}\left|\bfx^{\prime}\right|$,  we observe a stringy dipole,   
\be
\ba{ll}
\dis{\bfH\simeq G\,\frac{3\hat{\bfx}\left(\hat{\bfx}\bmcdot\bfS(t)\right)-\bfS(t)}{\left|\bfx\right|^{3}}\,,}\quad&\quad
\bfS(t)%=\half\!\dis{\int}\!\rd^{3}x~\bfx\bmtimes\bfK(t,\bfx)
=\dis{\int}\!\rd^{3}x~\bfs(t,\bfx)\,.
\ea
\ee

As an example, we consider a uniformly  `stringized' sphere of radius $a$,  with   constant $\rho$ and $\bfs$, to get (\textit{c.f.~}\cite{Jackson:1998nia})
\be
\ba{lll}
\Phis=\frac{4\pi}{3}\,\bfs\bmcdot\bfx\,,\quad&\quad\!\!
\bfH=\frac{8\pi G}{3}\,\bfs\quad&~~\mbox{for~}\left|\bfx\right|\leq a\,,\\
\Phis=\frac{4\pi a^{3}}{3}\,\frac{\bfs\bmcdot\bfx}{\,\left|\bfx\right|^{3}}\,,\quad&\quad\!\!
\bfH=\frac{4\pi G a^{3}}{3}\left(\frac{3\hat{\bfx}(\hat{\bfx}\bmcdot\bfs)-\bfs}{\left|\bfx\right|^{3}}\right)\quad&~~\mbox{for~}\left|\bfx\right|> a\,.
\ea
\ee
Thus, the total effective  mass density~(\ref{NewtonPot}) reads
\be
\rhoeff(t,\bfx)=\left\{\ba{ll}
\rho+\frac{\,16\pi G }{9}\left|\bfs\right|^{2}
\quad&\quad\mbox{for~}\left|\bfx\right|\leq a\\
\frac{\,4\pi G}{9}\left|\bfs\right|^{2}a^{6}\!\left(\frac{1+3\cos^{2}\!\theta}{\left|\bfx\right|^{6}}\right)
\quad&\quad\mbox{for~}\left|\bfx\right|>a\,,
\ea
\right.
\label{rhoeff}
\ee
where $\theta$ is the angle between $\bfs$ and $\bfx$. As anticipated from the general formula~(\ref{exact}), $\rhoeff$ has a $\left|\bfx\right|^{-6}$ profile or `halo'.

Another example is  of Dirac monopole type (singular),
\be
\ba{ll}
\dis{\bfB=Gq\int_{\bfx^{\prime}=\bf{0}}^{\infty}\,\rd\bfx^{\prime}\bmtimes
\frac{(\bfx-\bfx^{\prime})}{\,\left|\bfx-\bfx^{\prime}\right|^{3}}\,,}\quad&~~
\dis{\bfH=Gq\frac{\bfx}{~\left|\bfx\right|^{3}}\,,}
\ea
\ee
where the path should not cross the point of $\bfx$. The profile is now thicker   as  $\rhoeff=\frac{Gq^{2}}{4\pi}\left|\bfx\right|^{-4}$ which may be  comparable to some known    dark matter profiles~\cite{Hernquist:1990be,Navarro:1995iw}.  In fact,  this configuration corresponds to the  linearization of   a known  exact spherical solution to (\ref{REDFE})~\cite{Burgess:1994kq} which has been shown to feature a non-monotonic and hence non-Keplerian rotation curve~\cite{Ko:2016dxa}.

%%%%%%%%%%%%%%%%%%%%%%%%%%%%%%%%%%%%%%%%%%
%%%%%%%%%%%%%%%%%%%%%%%%%%%%%%%%%%%%%%%%%%
\vspace{5pt}
\section{Conclusion\label{SECDISCUSSION}}
\vspace{5pt}
We have shown that  the $\ODD$-completed General Relativity or Einstein Field Equations~(\ref{MASTER}) reduce in the non-relativistic limit to the Stringy  Newton Gravity~(\ref{MAIN}). 
 Symmetry-wise, the $\ODD$ of (\ref{MASTER}) is broken spontaneously  in (\ref{REDFE})  of which  General Covariance is  reduced to Lorentz symmetry in (\ref{lEDFE}) and further to Galilean symmetry in (\ref{MAIN}). It would be of interest to investigate whether    General Covariance can be recovered  as in  (stringy) Newton--Cartan Gravity~\cite{Andringa:2012uz,Hansen:2018ofj,Bergshoeff:2019pij}.
 
 The  final resulting formulas~(\ref{MAIN})   resemble  a hybrid of  Newton Gravity and Magnetostatics. Out of $D^{2}+1=17$ number of  components of the stringy energy-momentum tensor, $T_{AB}$~(\ref{MASTER})   or $\{K_{\mu\nu},\To\}$~(\ref{REDFE}), only four, \textit{i.e.~}the   mass density~$\rho$ and  the stringy current density~$\bfK$, participate in determining the Newton potential~(\ref{NewtonPot}). Different types of matter have different $\rho$ and $\bfK$ (or \textit{stringization} $\bfs$).  If the matter is  non-stringy  particle-like, $\bfK$ is trivial and we fully recover  Newton Gravity. On the other hand,  in the presence of distinct  kinds of matter, the center of $\rho$ may not  coincide with  that of $\bfH\bmcdot\bfH$. These might explain the  ghostly galaxies without dark matter (or without  $H$-flux sourced by $\bfK$)~\cite{ghost}  and  the Bullet Cluster~\cite{Clowe:2006eq} respectively.   It would be of the utmost interest to test    Stringy  Newton Gravity more rigorously against observations. For this, one needs to also   analyze light  or electromagnetism (for a recent discussion see \cite{Hansen:2019vqf}).  The $\ODD$ Symmetry Principle prescribes   gauge bosons  to   couple to  $g$ and $\phi$  but  not to $B$-field~\cite{Choi:2015bga},
\be
S_{{\rm{photon}}}=\dis{\int\rd^{4}x~-\quarter\sqrt{-g}e^{-2\phi}g^{\mu\rho}g^{\nu\sigma}F_{\mu\nu}F_{\rho\sigma}\,.}
\label{photon}
\ee
This seems to imply that a photon would not merely follow a null geodesic, but would  be also  influenced by  the dilaton~$\phi$, \textit{e.g.~}\cite{Basile:2019pic}.  Further analysis on the action~(\ref{photon}), using methods  such as Eikonal approximation~\cite{Eikonal}, is desirable. \\
~\\
\indent\textit{Acknowledgments.}   We wish to thank Stephen Angus, David Berman,  Gabriele Gionti, Chethan Krishnan, and  Jae-Weon Lee for helpful discussions.  This work  was  supported by  the National Research Foundation of Korea   through  the Grants,  NRF-2016R1D1A1B01015196  and NRF-2018H1D3A1A01030137~(Korea Research Fellowship Program).
\hfill

\end{document}